\documentclass{article}
\usepackage{amssymb}

\usepackage{graphicx}
\usepackage{amsmath}

\newtheorem{theorem}{Theorem}
\newtheorem{acknowledgement}[theorem]{Acknowledgement}

\input{tcilatex}

\begin{document}

\title{\textbf{Neutrinos in a vacuum dominated cosmology}}
\author{Manasse R. Mbonye \\
\textit{Michigan Center for Theoretical Physics,}\\
\textit{\ Physics Department, University of Michigan, Ann Arbor, Michigan
48109}}
\maketitle

\begin{abstract}
We explore the dynamics of neutrinos in a vacuum dominated cosmology. First
we show that such a geometry will induce a phase change in the eigenstates
of a massive neutrino and we calculate the phase change. We also calculate
the delay in the neutrino flight times in this geometry. Applying our
results to the presently observed background vacuum energy density, we find
that for neutrino sources further than $1.5\,Gpc$ away both effects become
non-trivial, being of the order of the standard relativistic corrections.
Such sources are within the obsevable Hubble Deep Field. The results which
are theoretically interesting are also potentially useful, in the future, as
detection techniques improve. For example such effects on neutrinos from
distant sources like supernovae could be used, in an independent method
alternative to standard candles, to constrain the dark energy density and
the deceleration parameter. The discussion is extended to investigate
Caianiello's inertial or maximal acceleration (MA) effects of such a vacuum
dominated spacetime on neutrino oscillations. Assuming that the MA
phenomenon exists, we find that its form as generated by the presently
observed vacuum energy density would still have little or no measurable
effect on neutrino phase evolution.
\end{abstract}

\smallskip

\subsection{Introduction}

In the last few years there have been two very interesting developments in
the field of physics. On the one hand, recent observations [1] strongly
suggest that the Hubble expansion does depart from that for a purely matter
dominated universe. The leading explanation is that the universe is
dominated by a mysterious low energy density vacuum $\rho _{\Lambda }\sim
\left( 1.6\times 10^{-3}eV\right) ^{4}$ whose dynamical effects are similar
to those of a cosmological constant $\Lambda $. On the other hand, recent
experiments at several research centers, such as the Super-Kamiokanda
Collaboration [2], have provided compelling evidence that the neutrino may
have a non-zero mass. Such evidence, coupled with the fact that they can
traverse very large distances unimpeded, makes neutrinos good long range
probes or good information carriers between any points cosmological
distances apart. In this paper we study the evolutionary consequences on the
propagation of massive neutrinos in the geometry of a vacuum dominated
cosmology. The results are used to discuss the effects of the observed [1]
background dark energy on neutrino dynamics.

The paper is arranged as follows. In section 2 we discuss neutrino
oscillations in a vacuum dominated cosmology. Section 3 deals with neutrino
flight time delay in this geometry. In section 4 we look for any neutrino
phase changes that may result from inertial effects. Section 5 concludes the
paper.

\subsection{Vacuum induced neutrino oscillations}

Neutrinos are produced and also detected through weak interactions. At their
production each particle emerges as flavor eigenstate $\left| \nu _{\alpha
}\right\rangle $. It is now widely believed [3] that each such state is a
coherent eigenstate of a linear superposition of mass eigenstates $\left|
\nu _{i}\right\rangle $ so that $\left| \nu _{\alpha }\right\rangle =%
\underset{i}{\sum }U_{\alpha i}^{\ast }\left| \nu _{i}\right\rangle $, where 
$U$ is unitary. These mass eigenstates propagate in spacetime as a plane
waves $\left| \nu _{i}(x,t)\right\rangle =\exp (-i\Phi _{i})\left| \nu
_{i}\right\rangle $, where $\Phi _{i}$ is the phase of the $i^{th}$ mass
eigenstate. Because the various mass eigenstates may have different energy
and momenta, they will propagate differently in space with the result that
their changing phases may interfere. This implies that a neutrino initially
produced at a spacetime point $P(t_{P},x_{P})$ with a flavor, $\nu _{\alpha
} $, in an eigenstate $\left| \nu _{\alpha }\right\rangle $ may have evolved
to a different flavor $\nu _{\beta }$, in an eigenstate $\left| \nu _{\beta
}\right\rangle $ by the time it is detected at a different spacetime point $%
Q(t_{Q},x_{Q})$. Such is the basis for neutrino oscillations. The
evolutionary effects on the relative phases of such mass eigenstates can be
driven by, among other things, the local geometry $g_{\mu \nu }$ of the
spacetime. Since the motion of given mass state, $\left| \nu
_{i}\right\rangle $, in a given geometry $g_{\mu \nu }$ is geodesic the
phase $\Phi _{i}$ can be written in a covariant form 
\begin{equation}
\Phi _{i}=\frac{1}{\hbar }\int p_{\mu }dx^{\mu }=\frac{1}{\hbar }\int g_{\mu
\nu }p^{\mu }dx^{\nu },  \tag{1}
\end{equation}
where $p_{\mu }$ is the conjugate momentum to $x^{\mu }$.

Recently several authors (see [4]\textbf{\ }citations therein) have
considered geometry effects on the phases of neutrinos propagating in the
gravitational field of a massive body. The current consensus is that in such
a geometry the resulting phase changes are negligibly small and also only
grow as $\ln r$, where $r$ is the distance traversed by the neutrino. In the
present discussion we consider the motion of neutrinos in a vacuum dominated
cosmology. In seeking for gravitationally induced phase changes arising
purely from the vacuum we assume that the latter does induce, on spacetime,
a de Sitter-type metric with an energy density $\rho _{\Lambda }\sim \left(
1.6\times 10^{-3}eV\right) ^{4}$. In comoving coordinates the line element
for such geometry is 
\begin{equation}
ds^{2}=-(1-\chi ^{2}r^{2})dt^{2}+(1-\chi ^{2}r^{2})^{-1}dr^{2}+r^{2}d\Omega
^{2},  \tag{2}
\end{equation}
where $\chi $ is related to the cosmological constant $\Lambda $ by $\Lambda
=3\chi ^{2}$. As this spacetime is isotropic we can, with no loss of
generality, restrict the motion to some fixed plane, say $\theta =\frac{\pi 
}{2}$. Thus, the phase $\Phi _{i}$ acquired by $\left| \nu _{i}\right\rangle 
$ propagating from point $P(t_{P},x_{P})$ to point $Q(t_{Q},x_{Q})$,\ in
such spacetime is 
\begin{equation}
\Phi _{i}=\frac{1}{\hbar }\int_{P}^{Q}\left( p_{t}dt+p_{r}dr+p_{\varphi
}d\varphi \right) .  \tag{3}
\end{equation}
Further, since this spacetime admits two Killing vectors $\partial _{t}$ and 
$\partial _{\varphi }$, the associated conjugate momenta $p_{t}$ and $%
p_{\varphi }$ are conserved quantities, so that 
\begin{equation}
p_{t}=-m_{i}(1-\chi ^{2}r^{2})\frac{dt}{ds}=-E=const.  \tag{4}
\end{equation}
and 
\begin{equation}
p_{\varphi }=m_{i}r^{2}\frac{d\varphi }{ds}=const.  \tag{5}
\end{equation}
Such quantities represent, respectively, the energy and the angular momentum
of the particle as seen by an observer in the region $r\rightarrow 0$. The
conjugate momenta components can be linked by the mass-shell relation, $%
g^{\mu \nu }p_{\mu }p_{\nu }-m_{i}^{2}=0$. Thus for the geometry under
consideration, we have 
\begin{equation}
-(1-\chi ^{2}r^{2})^{-1}\left( p_{t}\right) ^{2}+(1-\chi ^{2}r^{2})\left(
p_{r}\right) ^{2}+\frac{1}{r^{2}}\left( p_{\varphi }\right) ^{2}+m_{i}^{2}=0,
\tag{6}
\end{equation}
where 
\begin{equation}
p_{r}=m_{i}(1-\chi ^{2}r^{2})^{-1}\frac{dr}{ds}.  \tag{7}
\end{equation}

To keep the discussion simple we shall focus on the radial motion. In this
direction $d\varphi =0$ and the phase in (3) becomes

\begin{equation}
\Phi _{i(PQ)}=\frac{1}{\hbar }\int_{r_{P}}^{r_{Q}}\left( p_{r}-E\frac{dt}{dr}%
\right) dr,  \tag{8}
\end{equation}
where from (4) and (7) we have 
\begin{equation}
\frac{dt}{dr}=-(1-\chi ^{2}r^{2})^{-2}\frac{E}{p_{r}}.  \tag{9}
\end{equation}
Using (6) with $d\varphi =0$ one can express the radial component of
momentum $p_{r}$ in terms of the constant energy $E$. Equation (8) then
takes the form 
\begin{equation}
\Phi _{i(PQ)}=\frac{1}{\hbar }\int_{r_{P}}^{r_{Q}}\left( (1-\chi
^{2}r^{2})^{-1}\sqrt{E^{2}-m_{i}^{2}(1-\chi ^{2}r^{2})}-E\frac{dt}{dr}%
\right) dr.  \tag{10}
\end{equation}

As the neutrino propagates, the phases of the $i^{th}$\ and $j^{th}$ mass
states, say, will evolve differently so that at the detector $Q(t_{Q},x_{Q})$
a phase difference $\Delta \Phi =\Phi _{i(PQ)}-\Phi _{j(PQ)}=\Delta \Phi
_{r}+\Delta \Phi _{t}$ is observable as an interference pattern. Here and,
henceforth, $\Delta \Phi _{r}$ and $\Delta \Phi _{t}$ refer, respectively,
to the spatial-momentum and the temporal-energy contributions to the phase
difference. There are usually two different approaches used in calculating $%
\Delta \Phi $ and yield similar results.

In the one case one supposes that the neutrino mass eigenstates are
relativistic in their entire flight from source to detector [3]. This allows
a discussion of the motion in terms of geometric optics. In this
approximation such states evolve as plane waves propagating on a null
surface, $ds=0$. The particle energy $E$ can then be expanded in terms of $%
E_{null}$, the associated energy of the massless fields at the origin $(r=0)$
(and which is constant along the null trajectory) and $\Delta \Phi $
calculated by evaluating the integral in (10) along this null trajectory.
While the method yields the correct results it does so at the expense of
overshadowing (and seemingly countering) the physics in the central argument
that the neutrino is massive. The method also runs into problems if some
neutrinos may turn out to have significant mass. Nevertheless, its
simplicity is appealing.

On the other hand, one can take the view that neutrinos are, indeed, massive
and classically evolve the various eigenstates along their geodesics with
the energy $E$ and the conjugate momentum $p_{\varphi }$ as constants of
motion. This method, when applied to the case of a neutrino as a massive
particle with localized energy, appears counter-intuitive for the following
reason. If the various mass eigenstates with the same energy $E$ and
different radial momenta $p_{r}$ start at the same initial spacetime point $%
P(t_{P},x_{P})$, it is difficult to see how they could end up at the same
final spacetime point $Q(t_{Q},x_{Q})$ so they can interfere. The only way
such particles could interfere at $Q(t_{Q},x_{Q})$ is if they started at
different times $t_{P}$ and $t_{P}^{%
{\acute{}}%
}$ so that there is an initial time difference $\Delta t$ between their
points of origin. However this runs one into a further counter-intuitive
problem since in the first place the initial neutrino flavor $\nu _{\alpha }$
was supposed to be a localized particle.

Nevertheless, each of the above approximations yields the correct results
and we shall find it convenient in this treatment to utilize the latter
approach with modifications. Our arguments are related to those in [4] where
the authors consider a wave packet with a large flavor correlation length.
In our treatment, we assume that at the initial \ spacetime point $%
P(t_{P},x_{P})$ the mass states are produced with an energy width $W$
related to an energy spread $\Delta E$ about some average energy $\langle
E\rangle $ which classically turns out to be the constant energy $E$. This
spread then contributes a term $\frac{\Delta E}{W}$ to the phase difference, 
$\Delta \Phi $. Consequently, the masses $m_{i=1,2}$ will arrive at $%
Q(t_{Q},x_{Q})$ simultaneously and hence interfere provided the
temporal-energy contribution to the phase difference vanishes, i.e.

\begin{equation}
\Delta \Phi _{t}=\Delta \frac{1}{\hbar }\int_{P}^{Q}p_{t}dt-\frac{\Delta E}{W%
}=0.  \tag{11}
\end{equation}
With this we then have that, for interference to take place, the only active
contribution to the phase difference $\Delta \Phi $ is $\Delta \Phi _{r}$
given by 
\begin{equation}
\Delta \Phi _{r}=\Delta _{ij}\left[ \frac{1}{\hbar }\int_{P}^{Q}p_{r}dr%
\right] =\Delta _{12}\left[ \frac{1}{\hbar }\int_{r_{P}}^{r_{Q}}\left(
(1-\chi ^{2}r^{2})^{-1}\sqrt{E^{2}-m_{i=1,2}^{2}(1-\chi ^{2}r^{2})}\right) dr%
\right] ,  \tag{12}
\end{equation}
where $\Delta _{12}$ takes the difference between the two integrals
associated with the two different masses $m_{1}$ and $m_{2}$. Noting that $%
E^{2}<<m_{i}^{2}$ one finds, on reintroducing $c$ into (12), that 
\begin{equation}
\Delta \Phi _{r}=\frac{\Delta m^{2}c^{3}}{2\hbar E}\left( r_{Q}-r_{P}\right)
+\frac{1}{8}\frac{\Delta m^{4}c^{7}}{\hbar E^{3}}\left( r_{Q}-r_{P}\right) -%
\frac{1}{24}\frac{\Delta m^{4}c^{7}}{\hbar E^{3}}\chi ^{2}\left(
r_{Q}^{3}-r_{P}^{3}\right) +.....,  \tag{13}
\end{equation}
where $\Delta m^{2}=\left| m_{1}^{2}-m_{2}^{2}\right| $ and $\Delta
m^{4}=\left| m_{1}^{4}-m_{2}^{4}\right| $. The first two terms in (13) are
[4] the usual flat space $\Delta \Phi _{0}$ and the special relativistic
correction $\Delta \Phi _{rel}$, respectively, for neutrino oscillations.
The term in $\chi ^{2}$ is our result for the leading cosmological
background contribution $\Delta \Phi _{\Lambda }$ to the oscillation of
neutrinos. Notice that this cosmological term grows as $r^{3}$ while the
first two terms only grow as $r$. Clearly for neutrino sources at
cosmological distances $r\sim \frac{1}{\chi }$ the cosmological term can be
important. In particular, for the current estimates of the cosmological
constant at $\Lambda =3\chi ^{2}\approx 10^{-56}cm^{-2}$, the phase effects
due to the geometry on neutrinos from a source, like a supernova, some $%
1.5Gpc$\ away would be of the order of the special relativistic correction
term. Since $\Delta \Phi _{\Lambda }$ is opposite, in sign, to $\Delta \Phi
_{rel}$ then at such distances the relativistic corrections are
significantly suppressed so that $\Delta \Phi _{r}\longrightarrow \Delta
\Phi _{0}$. One can compare this result to that due to the gravitational
effect of a massive body. Bhattachrya et al [4] have shown that in this
latter geometry gravitational contributions to neutrino oscillations are
virtually negligible and grow only as $\ln \frac{r_{Q}}{r_{P}}$, where $%
r_{Q}<r_{P}$. As one can infer from our result, the effect due to geometry
under consideration evolves differently, becoming non-trivial at large
distances.

\subsection{Neutrino flight delay times}

One other possible effect due to the local geometry on the dynamics of a
massive neutrino is the flight time. Here we are interested in the time
delay between the flights of neutrinos and photons as induced by the
geometry of a vacuum dominated spacetime.\textit{\ }The neutrino flight time 
$\Delta t_{\nu }$ can be estimated from the preceding discussion.\ From (9)
we have that $\Delta t_{\nu }=-\int (1-\chi ^{2}r^{2})^{-2}\frac{E}{p_{r}}%
dr. $ Applying (4) and (7) gives 
\begin{equation}
\Delta t_{\nu }=-\int_{r_{P}}^{r_{q}}(1-\chi ^{2}r^{2})^{-1}\frac{E}{\sqrt{%
E^{2}-m_{i}^{2}(1-\chi ^{2}r^{2})}}dr.  \tag{14}
\end{equation}
Since $1>\chi ^{2}r^{2}$ and since for the neutrino we always have $%
E^{2}>>m_{i}^{2}(1-\chi ^{2}r^{2})$ we can expand each of the two terms of
the integrand. This gives, to order $\frac{m^{4}}{E^{4}}$ 
\begin{equation}
\Delta t_{\nu }=-\left( 1+\frac{m^{2}}{2E^{2}}+\frac{3m^{4}}{8E^{4}}\right)
\left( r_{Q}-r_{P}\right) -\frac{1}{3}\left( 1-\frac{3m^{4}}{8E^{4}}\right)
\chi ^{2}\left( r_{Q}^{3}-r_{P}^{3}\right) +....  \tag{15}
\end{equation}
The first term is the flight time in a Minkowski space time. It contains the
classical and higher order relativistic contributions. \ The second term $%
\sim r^{3}$ results from the modifications of the spacetime geometry by the
cosmological constant.

On the other hand the flight time for a photon leaving the same spacetime
point $P(t_{P},x_{P})$ as the neutrino to the same detector $Q(t_{Q},x_{Q})$
can be obtained by integrating $\left( \frac{dt}{dr}\right) _{null}$ along
the null trajectory. \ We have from (9) that $\left( \frac{dt}{dr}\right)
_{null}=-\left[ (1-\chi ^{2}r^{2})^{-2}\frac{E}{p_{r}}\right]
_{p_{r}=g_{tt}E}$. Then the photon flight time $\Delta t_{\gamma }$ is given
by 
\begin{equation}
\Delta t_{\gamma }=\int \left( \frac{dt}{dr}\right) _{null}dr=-\int (1-\chi
^{2}r^{2})^{-1}dr.  \tag{16}
\end{equation}
which evaluates to $\Delta t_{\gamma }=-\frac{1}{2\chi }\ln \left[ \frac{%
1+\chi r}{1-\chi r}\right] _{r_{P}}^{r_{Q}}$ . This result can be expanded
for $r<\frac{1}{\chi }$ to give

\begin{equation}
\Delta t_{\gamma }=-\left( r_{Q}-r_{P}\right) -\frac{1}{3}\chi ^{2}\left(
r_{Q}^{3}-r_{P}^{3}\right) -\frac{1}{5}\chi ^{4}\left(
r_{Q}^{5}-r_{P}^{5}\right) +...  \tag{17}
\end{equation}
The delay $\Delta t_{\nu \gamma }$\ in neutrino arrival time with respect to
photon arrival time is then given by $\Delta t_{\nu \gamma }=\Delta t_{\nu
}-\Delta t_{\gamma }$. From equations (15) and (17) we find (on
re-introducing $c^{\prime }s$) that 
\begin{equation}
\Delta t_{\nu \gamma }=-\left( \frac{m^{2}c^{3}}{2E^{2}}+\frac{3m^{4}c^{7}}{%
8E^{4}}\right) \left( r_{Q}-r_{P}\right) +\left( \frac{m^{4}c^{7}}{8E^{4}}%
\right) \chi ^{2}\left( r_{Q}^{3}-r_{P}^{3}\right) +...  \tag{18}
\end{equation}
Our result departs from the usual result in flat space by the term 
\begin{equation}
\delta t_{\nu \gamma }=\left( \frac{m^{4}c^{7}}{8E^{4}}\right) \chi
^{2}\left( r_{Q}^{3}-r_{P}^{3}\right)  \tag{19}
\end{equation}
resulting from the vacuum effects.. We note that the only adjustable
parameter in this result is the distance from the source to the detector. As
in the phase change $\Delta \Phi _{r}$\ result in (13) we find that for the
current estimates of the cosmological constant at $\Lambda =3\chi
^{2}\approx 5\times 10^{-56}cm^{-2}$, the time delay effects due to the
geometry on massive neutrinos from a source, like a supernova, some $1.5Gpc$%
\ away would be of the order of the relativistic correction term. The
result, then, is that for neutrino sources at distances $r\gtrsim 1.5\,Gpc$
the relativistic corrections are, again, significantly suppressed by the
(opposite sign) corrections $\delta t_{\nu \gamma }$ due to the
vacuum-induced geometry.

\subsection{Vacuum induced inertial effects and neutrino phases}

As a final consideration on gravitationally induced phase changes in
neutrino eigenstates we take a look at the possible effects originating from
a phenomenon that has lately been common in the literature, namely that of
maximal acceleration (MA) of particles. We should mention that, to our
knowledge, such a phenomenon has not yet been observed. In order to set the
problem we find it useful to lay out the background. The geometry
experienced by a particle of mass $m$ accelerated in a background spacetime $%
g_{\mu \nu }$ was first discussed by Caianiello [5]\textbf{. }According to
Caianiello\textbf{\ }such a geometry is defined [5] on an eight-dimensional
manifold $\mathbf{M}_{8}$ by a metric $d\tilde{s}^{2}=$ $g_{AB}dX^{A}dX^{B}$%
, where $\left( A=0,1,2...7\right) $ and $X^{A}=\left( x^{\mu },\frac{c^{2}}{%
\mathcal{A}_{m}}\frac{d\dot{x}^{\mu }}{ds}\right) $. Here $g_{AB}=g_{\mu \nu
}\otimes g_{\mu \nu }$, $ds$ is the usual four-dimensional element given by $%
ds^{2}=$ $g_{\mu \nu }dx^{\mu }dx^{\mu }$, with $\mu =0,1..3$ and $\mathcal{A%
}_{m}$ is called the maximal acceleration of the particle mass $m$, given by 
$\mathcal{A}_{m}=\frac{2mc^{3}}{\hbar }$. An effective four-dimensional
spacetime that takes consideration of the maximal acceleration of the
particle [6]\textbf{\ }can be defined as an imbedding in $\mathbf{M}_{8}$.
The metric $\tilde{g}_{\mu \nu }$ induced on such a hyperface imbedded in $%
\mathbf{M}_{8}$ gives rise to a line element 
\begin{equation}
d\tilde{s}^{2}=\sigma ^{2}\left( x\right) g_{\alpha \beta }dx^{\alpha
}dx^{\beta },  \tag{20}
\end{equation}
where the conformal factor $\sigma ^{2}\left( x\right) $ is given by $\sigma
^{2}\left( x\right) =1+g_{\mu \nu }\frac{c^{4}}{\mathcal{A}_{m}^{2}}\ddot{x}%
^{\mu }\ddot{x}^{\nu }$. The appearance of the quantity $\mathcal{A}_{m}=%
\frac{2mc^{3}}{\hbar }$, where $m$ is the rest mass of the particle, implies
that the geodesics are mass dependant, in violation of the equivalence
principle. Moreover, the accelerations $\frac{d^{2}x^{\mu }}{ds^{2}}=\ddot{x}%
^{\mu }$ which are related to the Newtonian force are not covariant
quantities and, with respect to symmetries, the conformal factor $\sigma
^{2} $ is neither invariant nor can it be removed by general coordinate
transformations [7].\ The metric in (20) does therefore not satisfy the
standard requirements of general relativity. The original derivations [5,6]\
which aimed at relating quantum mechanics to gravity use special relativity
in flat Minkowiski $\left( g_{\mu \nu }\rightarrow \eta _{\mu \nu }\right) $%
\ spacetime. One then assumes that the technique should yield reasonably
accurate results, at least in a locally flat environment as generated by
weak gravitational fields.

On a cosmological scale, such a weak field environment can be provided by
the observed [1] background vacuum energy density in the form of a
cosmological constant $\Lambda =3\chi ^{2}$, and whose metric is given by
(2). We shall presently derive, based on (20), expressions for MA in a
vacuum dominated cosmology and use such results to seek for any associated
effects on the evolution of neutrino phases.

The symmetry of this spacetime leads to\textit{\ }$\sigma ^{2}=\sigma
^{2}\left( r,\theta \right) $. Accordingly a relativistic particle, like a
neutrino mass $m$, moving in such a cosmological environment would
experience a geometry

\begin{equation}
d\tilde{s}^{2}=\sigma ^{2}\left( r,\theta \right) \left[ -(1-\chi
^{2}r^{2})dt^{2}+(1-\chi ^{2}r^{2})^{-1}dr^{2}+r^{2}d\Omega ^{2}\right] . 
\tag{21}
\end{equation}
One can restrict the motion to some fixed plane, say $\theta =\frac{\pi }{2}$%
. We then have that , 
\begin{equation}
\sigma ^{2}=\sigma ^{2}\left( r\right) =1+\frac{c^{4}}{\mathcal{A}_{m}^{2}}%
\left[ \left( 1-\chi ^{2}r^{2}\right) \left( \frac{d^{2}t}{ds^{2}}\right)
^{2}-(1-\chi ^{2}r^{2})^{-1}\left( \frac{d^{2}r}{ds^{2}}\right)
^{2}-r^{2}\left( \frac{d^{2}\varphi }{ds^{2}}\right) ^{2}\right] .  \tag{22}
\end{equation}
The quantities $\ddot{x}^{\mu }=\frac{d^{2}x^{\mu }}{ds^{2}}$ can be written
down in terms of the total energy $E$ and the angular momentum $L$. Using
the mass-shell relation (6) and the energy conservation equations (4) and
(5) we find that 
\begin{eqnarray}
\ddot{t}^{2} &=&\left( \frac{E}{m}\right) ^{2}\frac{4\chi ^{4}r^{2}}{\left(
1-\chi ^{2}r^{2}\right) ^{4}}\left[ \left( \frac{E}{m}\right) ^{2}-\left(
1-\chi ^{2}r^{2}\right) \left( 1+\frac{L^{2}}{mr^{2}}\right) \right] , 
\notag \\
&&  \notag \\
\ddot{r}^{2} &=&\left[ \chi ^{2}r+\frac{L^{2}}{mr^{3}}\right] ^{2}, 
\TCItag{23} \\
&&  \notag \\
\ddot{\varphi}^{2} &=&\frac{L^{2}}{mr^{6}}\left[ \left( \frac{E}{m}\right)
^{2}-\left( 1-\chi ^{2}r^{2}\right) \left( 1+\frac{L^{2}}{mr^{2}}\right) %
\right] .  \notag
\end{eqnarray}
Equations (22) and (23) give the conformal factor in a de Sitter spacetime
as 
\begin{equation}
\sigma ^{2}\left( r\right) =1+\frac{c^{4}}{\mathcal{A}_{m}^{2}}\left[ 
\begin{array}{c}
\left( \frac{E}{m}\right) ^{2}\frac{4\chi ^{4}r^{2}}{\left( 1-\chi
^{2}r^{2}\right) ^{3}}\left[ \left( \frac{E}{m}\right) ^{2}-\left( 1-\chi
^{2}r^{2}\right) \left( 1+\frac{L^{2}}{mr^{2}}\right) \right] \\ 
-(1-\chi ^{2}r^{2})^{-1}\left[ \chi ^{2}r+\frac{L^{2}}{mr^{3}}\right] ^{2}
\\ 
-\frac{L^{2}}{mr^{4}}\left[ \left( \frac{E}{m}\right) ^{2}-\left( 1-\chi
^{2}r^{2}\right) \left( 1+\frac{L^{2}}{mr^{2}}\right) \right] .
\end{array}
\right] .  \tag{24}
\end{equation}

As has been our approach we shall consider here, for the neutrino mass
eigenstate $\left| \nu _{i}\right\rangle $, only the radial motion $\left(
d\varphi =0\right) $. Then the effective conformal factor becomes 
\begin{equation}
\sigma ^{2}\left( r\right) =1+\frac{c^{4}}{\mathcal{A}_{m}^{2}}\left[ \frac{%
\chi ^{4}r^{2}}{\left( 1-\chi ^{2}r^{2}\right) }\right] \left\{ \left( \frac{%
E}{m}\right) ^{2}\frac{4}{\left( 1-\chi ^{2}r^{2}\right) ^{2}}\left[ \left( 
\frac{E}{m}\right) ^{2}-\left( 1-\chi ^{2}r^{2}\right) \right] -1\right\} . 
\tag{25}
\end{equation}
Following (1), the phase $\tilde{\Phi}_{i}$ induced on a neutrino mass
eigenstate $\left| \nu _{i}\right\rangle $ propagating in this geometry is
now given by $\tilde{\Phi}_{i}=\frac{1}{\hbar }\int p_{\mu }dx^{\mu }$,
where $\tilde{p}_{\mu }=m\tilde{g}_{\mu \nu }\frac{dx^{\nu }}{d\tilde{s}}$
is the four-momentum and $\tilde{g}_{\mu \nu }=\sigma ^{2}\left( r\right)
g_{\mu \nu }$. The mass-shell condition (3) is modified to $\tilde{g}^{\mu
\nu }p_{\mu }p_{\nu }-m_{i}^{2}=0$ and yields $\tilde{p}_{r}=(1-\chi
^{2}r^{2})^{-1}\sqrt{E^{2}-m_{i}^{2}\sigma ^{2}\left( r\right) (1-\chi
^{2}r^{2})}$. As a result, one can now address the conditions for
interference to take place at the detector $Q$. Applying the same arguments
leading to (12) one finds that the only active contribution to the phase
difference $\Delta \tilde{\Phi}$ is $\Delta \tilde{\Phi}_{r}$ given by 
\begin{equation}
\Delta \tilde{\Phi}_{r}=\tilde{\Delta}_{ij}\left[ \int_{P}^{Q}\tilde{p}_{r}dr%
\right] =\tilde{\Delta}_{12}\left[ \int_{r_{P}}^{r_{Q}}\left( (1-\chi
^{2}r^{2})^{-1}\sqrt{E^{2}-m_{i=1,2}^{2}\sigma ^{2}\left( r\right) (1-\chi
^{2}r^{2})}\right) dr\right] ,  \tag{26}
\end{equation}
where $\sigma ^{2}$ is given by (25) and where $\Delta _{12}$ takes the
difference between the two integrals associated with the two different
masses $m_{1}$ and $m_{2}$. Equation (26) can be evaluated to give $\Delta 
\tilde{\Phi}_{r}=\Delta \Phi _{r}+\Delta \Phi _{\sigma \left( r\right) }$,
where $\Delta \Phi _{r}$ is given by (13) and $\Delta \Phi _{\sigma \left(
r\right) }$ is the new contribution involving the MA term and is given (up
to terms in $r^{3}$) by.

\begin{equation}
\Delta \Phi _{\sigma \left( r\right) }=-\frac{\hbar }{6c^{3}}\left[ \frac{%
E^{3}}{c^{4}}\frac{\Delta m^{4}}{\left( m_{1}m_{2}\right) ^{4}}-E\frac{%
\Delta m^{2}}{2\left( m_{1}m_{2}\right) ^{2}}\right] \chi ^{4}\left(
r_{Q}^{3}-r_{P}^{3}\right) .  \tag{27}
\end{equation}
Here we have, again, restored all $c^{\prime }s$ to facilitate numerical
estimates. Clearly $\Delta \Phi _{\sigma \left( r\right) }$\ contributes
differently to $\Delta \tilde{\Phi}_{r}$\ than the regular $\Delta \Phi _{r}$
given\ in (13). One notices from equation (27) that the leading term in $%
\Delta \Phi _{\sigma \left( r\right) }$\ is second order in $\chi ^{2}$ and
at the same time proportional to $E^{3}$. If we compare this leading MA term
in (27) to the relativistic term $\Delta \Phi _{rel}=\frac{1}{8}\frac{\Delta
m^{4}c^{7}}{\hbar E^{3}}\left( r_{Q}-r_{P}\right) $ in (13) we see that even
at cosmological distances\textit{\ }$r\sim \frac{1}{\chi }$ the ratio $\frac{%
\Delta \Phi _{\sigma \left( r\right) }}{\Delta \Phi _{rel}}$ is controlled
by the cosmological constant,

\begin{equation}
\frac{\Delta \Phi _{\sigma \left( r\right) }}{\Delta \Phi _{rel}}\mathit{\ }%
\approx \frac{4}{3}\left[ \left( \hbar c\right) ^{2}\frac{E^{6}}{\left(
mc^{2}\right) ^{8}}\right] \chi ^{2}.  \tag{28}
\end{equation}

For supernova neutrinos, say, with energy $E\sim 10$\thinspace $eV$ and mass 
$mc^{2}\sim 0.1\,\,eV$ one can deduce that the above ratio is extremely
small, $\sim 10^{-3}\chi ^{2}$. Evidently $\frac{\Delta \Phi _{\sigma \left(
r\right) }}{\Delta \Phi _{rel}}$ becomes even smaller for $r<<\frac{1}{\chi }
$, dimnishing (see (27)) as $\chi ^{4}r^{3}$. Consequently, assuming the MA
effect exists, its cosmological form still appears to have little or no
measurable contribution to neutrino oscillations. On the other hand it has
recently been shown [8] that in a Schwarzschild geometry $\Delta \Phi
_{\sigma \left( r\right) }$ can make significant contributions to $\Delta
\Phi $.

\subsection{Conclusion}

In conclusion, we have investigated the gravitational effects of vacuum
energy on the propagation of neutrinos. To isolate such effects, we assumed
the vacuum defines on spacetime a de Sitter-type geometry (with a positive
cosmological constant). It is found that such vacuum geometry induces a
phase change $\Delta \Phi _{\Lambda }$ in the neutrino eigenstates. This
phase change grows as $r^{3}$, where $r$ is the distance of the source from
the detector. We have also calculated the neutrino delay time induced by
such a geometry and found a similar cubic growth in the radial component of
the motion. In particular, we find that for $r\gtrsim \frac{1}{\chi }$ the
phase change $\Delta \Phi _{\Lambda }$ contribution to $\Delta \Phi $ and
the flight time delay $\delta t_{\nu \gamma }$ contribution to $\Delta
t_{\nu \gamma }$ can both be of the order of their respective special
relativistic contributions. Applying our results to background vacuum energy
density associated with [1] the presently observed $\Lambda \sim 5\times
10^{-56}$\thinspace $cm^{-2}$, we find that for neutrino sources further
than $1.5\,Gpc$ away both the above effects become non-trivial. Such sources
are well within the Hubble Deep Field. The results which are theoretically
interesting are also potentially useful, in the future, as detection
techniques improve. For example such effects, on neutrinos from distant
sources like supernovae, could be used in an independent method alternative
to standard candles, to constrain the back ground dark energy density and
the deceleration parameter. Undoubtedly, making use of such information
depends on improved future techniques to record events from weak neutrino
fluxes as those originating from such sources cosmological distances away.

Finally, the discussion was extended to investigate Caianiello's inertial or
maximal acceleration (MA) effects of such a vacuum dominated spacetime on
neutrino oscillations. Assuming that the MA phenomenon exists, we find that
its form as generated by the presently observed $\Lambda \sim 10^{-56}$%
\thinspace $cm^{-2}$ would still have little or no measurable effect on
neutrino phase evolution, even at cosmological distances, $r\sim \frac{1}{%
\chi }$.

\begin{acknowledgement}
We would like to thank Fred Adams and Ronald Mallett for some useful
discussions and Greg Tarle for originally posing the problem to the author.
\end{acknowledgement}

\begin{acknowledgement}
This work was supported with funds from the University of Michigan.
\end{acknowledgement}

\end{document}